\documentclass[twoside]{article}

\usepackage{lipsum} 

\usepackage[sc]{mathpazo} 
\usepackage[T1]{fontenc} 
\usepackage{microtype} 

\usepackage[hmarginratio=1:1,top=32mm,columnsep=20pt]{geometry} 
\usepackage{multicol} 
\usepackage[hang, small,labelfont=bf,up,textfont=it,up]{caption} 
\usepackage{booktabs} 
\usepackage{float} 
\usepackage{hyperref} 
\usepackage{natbib} 
\usepackage{graphicx}
\usepackage{caption} 
\usepackage{subcaption} 
\usepackage{lettrine} 
\usepackage{paralist} 

\usepackage{abstract} 

\usepackage{titlesec} 
\renewcommand\thesection{\Roman{section}} 
\renewcommand\thesubsection{\Roman{subsection}} 
\titleformat{\section}[block]{\large\scshape\centering}{\thesection.}{1em}{} 
\titleformat{\subsection}[block]{\large}{\thesubsection.}{1em}{} 

\usepackage{fancyhdr} 
\title{\vspace{-15mm}\fontsize{24pt}{10pt}\selectfont\textbf{Astrology in the Era of Exoplanets}} 

\author{
\large
\textsc{Michael B. Lund}\\
\normalsize Vanderbilt University \\ 
\normalsize \href{mailto:michael.b.lund@vanderbilt.edu}{michael.b.lund@vanderbilt.edu} 
\vspace{-5mm}
}
\date{}

\begin{document}

\maketitle 

\thispagestyle{fancy} 


\begin{abstract}

\noindent The last two decades have seen the number of known exoplanets increase from a small handful to nearly 2000 known exoplanets, thousands more planet candidates, and several upcoming missions that are expected to further increase the population of known exoplanets. Beyond the strictly scientific questions that this has led to regarding planet formation and frequency, this has also led to broader questions such as the philosophical implications of life elsewhere in the universe and the future of human civilization and space exploration. One additional realm that hasn't been adequately considered, however, is that this large increase in exoplanets would also impact claims regarding astrology. In this paper we look at the distribution of planets across the sky and along the Ecliptic, as well as the current and future implications of this planet distribution.

\end{abstract}


\begin{multicols}{2} 

\section{Introduction}
\lettrine[nindent=0em,lines=3]{T} he history of astronomy has been intricately tied with the history of astrology. The origins of the constellations can be traced back at least as far as how the Sumerians divided the sky into what we now know as the Zodiac \citep{Verderame2009}. The usage of astrological artifacts in the area around the Mediterranean Sea, such as an ivory astrologer's board, date back over 2000 years \citep{Forenbaher2011}. Humanity's perspective of the universe has changed over the last millenia, however astrology has lingered on. A system that relies on the positions of the planets, astrology has endured and adapted as our understanding has grown, from the transition to a heliocentric system \citep{Copernicus1543}, to the discoveries of Uranus \citep{Herschel1781}, Neptune \citep{Galle1846, Airy1846}, and Pluto \citep{Shapley1930}.

Within astrology, all planets have approximatly equal significance, despite a wide range of masses and distances; this rules out any of the fundamental forces as we understand them for being an explanation of how astrology influences the Earth. Quantum mechanics, however, has introduced unconventional physical behavior that can result in effects being transmitted at large distances, such as the unusual behavior of quantum entanglement \citep{Herbst2015}. Astrology's ability to never seem wrong, despite the changing understanding of the cosmos, also can be explained through the lens of quantum mechanics. At first glance, astrology's inability to predict undiscovered planets through imperfect forcasts seems to invalidate astrology as a study, but this can be explained if planets only have influence if they have been observed first. This may seem too fantastic to be true, but it can also be viewed simply as another example of the observation effect, in the same manner as how particles behave as a probability wave until observations cause this wave to collapse \citep{Buks1998}. At this point of first discovery for planets, the range of probable effects collapses to a single set of effects that astrology then incorporates.

That astrology is independent of the distance to planets is the key to understanding the most significant change to astrology, possibly in its entire history. From time immemorial, the number of known planets under any definition has never exceeded more than 10-20 bodies. In the last 20 years, however, this number has started to increase greatly as planets have been discovered outside our own solar system. With distance posing no obsticle to astrology, these planets need to be incorporated into any reasonable astrological framework. In this paper, we examine how the distribution of exoplanets in the sky has shifted the cosmic balance of humanity's fate.
\begin{figure*}[!htb]
  \begin{center}
   \includegraphics[width=\textwidth]{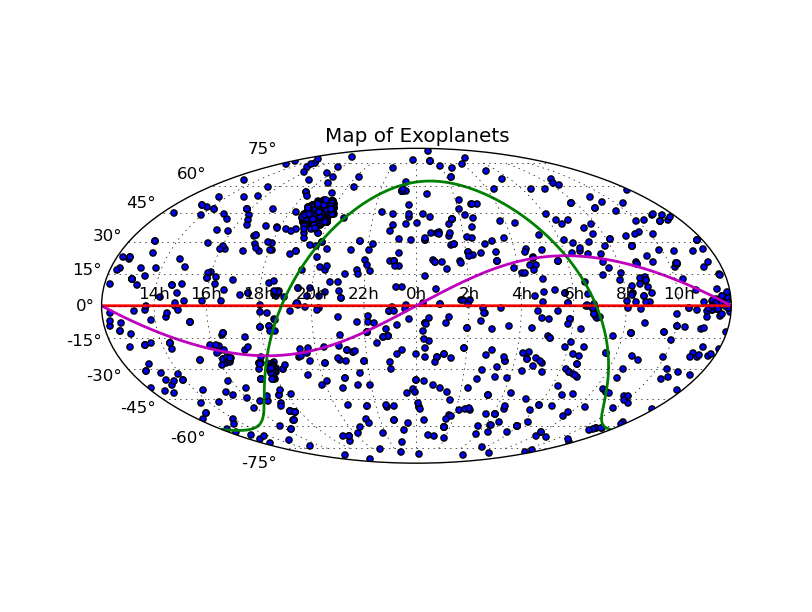}
  \end{center}
  \caption{A map of all known exoplanets. The red line marks the celestial equator, the magenta line marks the Ecliptic, and the green line marks the plane of the Milky Way. The highest density of known transiting planets is the location of the \emph{Kepler} field.}
  \label{fig:map}
\end{figure*}

\section{Mapping Exoplanets}
For this paper, we have  used the NASA Exoplanet Archive's\footnote{http://exoplanetarchive.ipac.caltech.edu/} database to gather a list of all known exoplanets, as well as the RA, DEC, and method of detection for each planet. As traditional astrology has treated each planet seperately, we treat each exoplanet in a system independently as well. In Fig~\ref{fig:map} we map the distribution of exoplanets on the sky.

\begin{table}[H]
\caption{Constellations and Properties}
\label{table:const}
\centering
\begin{tabular}{lll}
\toprule
Constellation & Element & Quality \\
\midrule
Aries & Fire & Cardinal \\
Taurus & Earth & Fixed \\
Gemini & Air & Mutable \\
Cancer & Water & Cardinal \\
Leo & Fire & Fixed \\
Virgo & Earth & Mutable \\
Libra & Air & Cardinal \\
Scorpio & Water & Fixed \\
Sagittarius & Fire & Mutable \\
Capricorn & Earth & Cardinal \\
Aquarius & Air & Fixed \\
Pisces & Water & Mutable \\
\bottomrule
\end{tabular}
\end{table}

\begin{figure*}[!htb]
  \begin{center}
   \includegraphics[width=\textwidth]{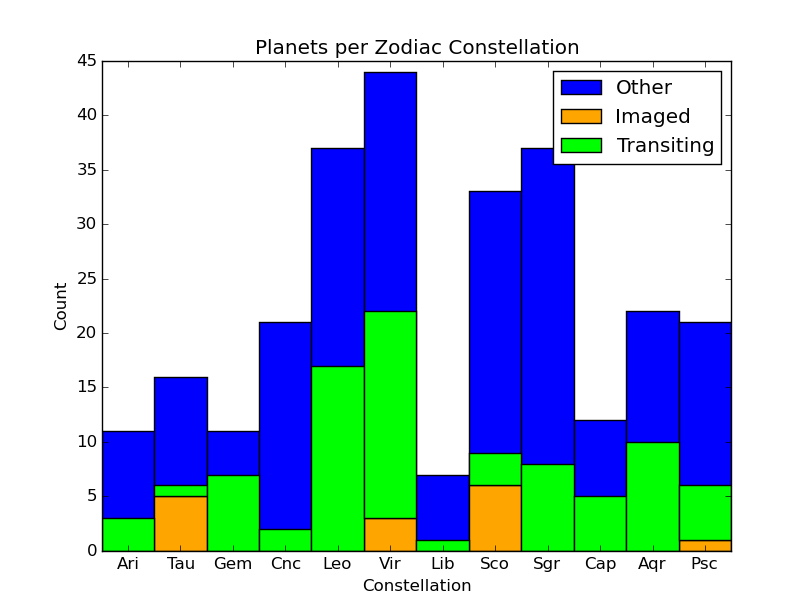}
  \end{center}
  \caption{The number of exoplanets in each constellation in the Zodiac. Planets that have been detected by direct imaging are in orange, planets that have been directed through transits are in lime, and planets detected using any other methods (e.g. Transit Timing Variation, Microlensing, etc) are in blue.}
  \label{fig:bar}
\end{figure*}

\begin{figure*}[!htb]
  \begin{center}
    \begin{subfigure}[b]{0.46\textwidth}
      \includegraphics[width=\textwidth]{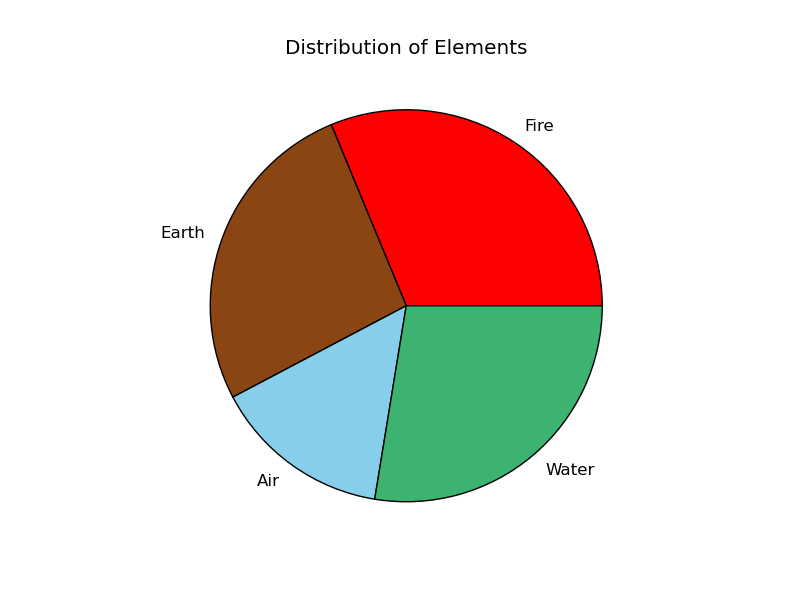}
    \end{subfigure}
    \begin{subfigure}[b]{0.46\textwidth}
      \includegraphics[width=\textwidth]{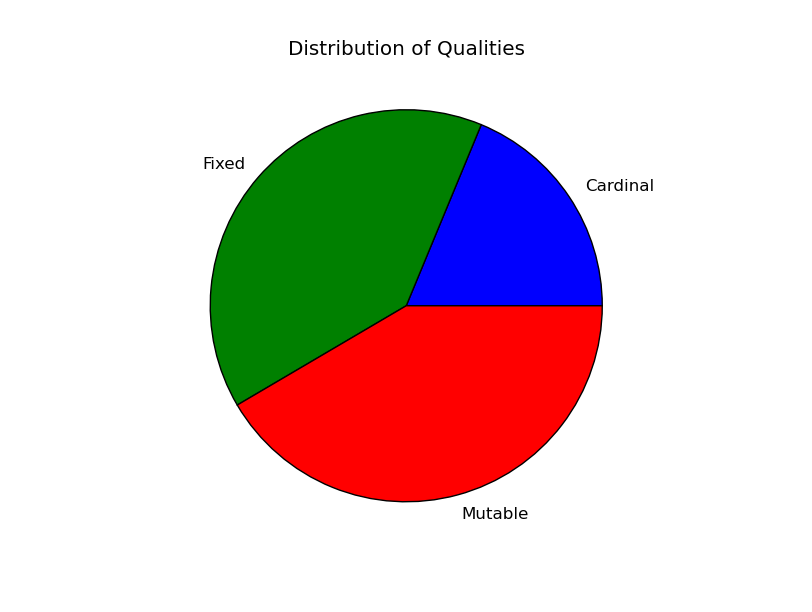}
    \end{subfigure}
  \end{center}
  \caption{On the left is the distribution of constellations based off of the element each constellation is associated with. On the right is the distribution based off of the quality of each constellation.}
  \label{fig:pie}
\end{figure*}

We have used the PyEphem\footnote{http://rhodesmill.org/pyephem/} package in Python to calculate the constellation that every exoplanet appears in. While exoplanets are distributed throughout the sky, we are only concerned here with those that occur along the Ecliptic. To that end, we only focus on planets that fall into one of the constellations of the classical Zodiac, as is shown in the distribution in Fig~\ref{fig:bar}. While the case can be made that just as planets within our own Solar System only entered astrology on optical detection, the case could also be made that exoplanets only are observed when they have been directly imaged, and the subtler point that transiting planets may also count as being visually observed, but we consider that question to be outside the scope of this paper. Finally, each of the constellations is also associated with two additional properties, the Element and the Quality (Table~\ref{table:const}. These distributions are included in Fig~\ref{fig:pie}. In each of these cases, a disproportionate number of planets in these categories represents an absence or overabundance of the associated traits.

\section{Discussion}
As a resource for the traits associated with various signs and properties, we use as an online resource\footnote{http://www.bobmarksastrologer.com/}. The first analysis is simply the distribution by constellation. The first immediate result is the paucity of planets in the constellation Libra. Described as representing "manners and the social graces, things which make cooperation and civil society possible," the rise in planets in the other signs also would represent a comparitive decline in these qualities. It is also worth noting that the number of planets that have been discovered has sharply increased in the last 10 years, and this is also a time frame where political polarization has greatly increased in the United States \footnote{http://www.people-press.org/2014/06/12/political-polarization-in-the-american-public/}, and numerous ongoing civil wars and other conflicts have begun in the last several years worldwide. In contrast, the largest share of planets appear in Virgo, a constellation associated with "always hav[ing] to keep busy" and is linked to criticism and self-critism. Again a reflection of social concerns of the last decade, the increasing presence of technology such as smartphones that have increased connectivity has also been associated with leisure time being intruded upon by work as employees are expected to always be in contact. At the same time, concerns with increasing depression and anxiety can reflect the dangerous effects of self-criticism when it goes unchecked\footnote{http://www.nytimes.com/2014/03/25/opinion/a-great-depression.html}.

Moving on to our analysis of the distribution of planets by element, the largest share of planets are in the Fire signs of Aries, Leo, and Sagittarius. This group of signs are associated with someone being more self-centered, and while this can be positively expressed as confidence and inspiration, this can also be negatively expressed as "a total lack of empathy, the inability to believe that other people are alive and have rights." Recent work has shown concerning results that empathy for others is on the decline, including in both public polling\footnote{http://www.theguardian.com/culture/australia-culture-blog/2014/feb/26/is-australia-losing-its-empathy} and in sociological studies \citep{Konrath2010}. As for the qualities of the signs, Fixed and Mutable signs each represent nearly half of the exoplanets. The Fixed signs represent determination and persistence, but this can often be expressed as the need to be right, and fighting for one's beliefs, regardless of any contrary beliefs. Again, this echoes the issues discussed above from a lack of cooperation and an increase in polarization. The Mutable signs are considered to be the communicators, and will only stay focused on something so long as interest is being maintained, and then move on to something new. The large increase of exoplanets in Mutable signs over the last two decades also corresponds with the development of a society increasingly dependent on social media worldwide\footnote{http://www.statista.com/statistics/278414/number-of-worldwide-social-network-users/}, and shifts from social media site to social media site as new trends are developed (e.g. Friendster to MySpace to Facebook).

\section{Summary}

Exoplanets have gone unaddressed in astrology, however there is no reason why astrology should simply cut off at the edge of our Solar System, and as astrology has had to dramatically adjust in the past with new astronomical discoveries, it would be too arbitrary to decide to not do that now. We have looked at the distribution of known exoplanets along the constellations of the zodiac, and examined how this correlates with observed sociological trends in the last 2 decades. As additional exoplanets are found, astrological interpretations would suggest that these global trends should continue to shift. This would further suggest that astronomy is in a unique position to improve humanity through targeted exoplanet searches. In particular, the greatest opportunity for establishing more cooperation and a more civil society would be to focus on searching Libra for additional planets. The globular cluster NGC 5897 would present one large population of stars that could provide a source of planets to alter our fate. Recent work carried out within our solar system also may result in additional planets within our own solar system that will need to be incorporated into astrology, however these challenges are well beyond the scope we have discussed here.


\bibliographystyle{apalike}
\bibliography{library}


\end{multicols}

\end{document}